\documentclass[aps,prl,superscriptaddress,amsmath,amssymb,twocolumn,fleqn,nofootinbib,preprintnumbers]{revtex4-1}
\usepackage{xcolor}
\usepackage{graphicx}
\usepackage{colordvi}
\usepackage{mathtools}
\usepackage{empheq}
\usepackage{bm}
\usepackage{comment}
\usepackage{floatrow}
\usepackage{braket}
\usepackage{soul}
\usepackage{siunitx}
\DeclareSIUnit \parsec {pc}
\usepackage{amssymb}
\usepackage{amsmath}
\usepackage{enumerate}
\usepackage{hyperref}
\def\be{\begin{equation}}
	\def\ee{\end{equation}}
\def\ba{\begin{eqnarray}}
	\def\ea{\end{eqnarray}}

\newcommand{\dd}{\mathrm{d}}
\def\l{\left}
\def\r{\right}
\def\f{\frac}
\DeclareMathOperator\erf{Erf}
\usepackage{soul}

\begin{document}

	\title{Quest for CMB spectral distortions to probe the scalar-induced gravitational wave background interpretation of pulsar timing array data}
	
     \author{Matteo Tagliazucchi}\email{matteo.tagliazucchi2@unibo.it}
	\affiliation{Dipartimento di Fisica e Astronomia, Alma Mater Studiorum
		Universit\`a di Bologna, \\
		Via Gobetti, 93/2, I-40129 Bologna, Italy}
  	\affiliation{INAF/OAS Bologna, via Gobetti 101, I-40129 Bologna, Italy}
	
	\author{Matteo Braglia}\email{mb9289@nyu.edu}
	\affiliation{Center for Cosmology and Particle Physics, New York University, 726 Broadway, New York, NY 10003, USA}
	\affiliation{INAF/OAS Bologna, via Gobetti 101, I-40129 Bologna, Italy}	
	
	\author{Fabio Finelli}\email{fabio.finelli@inaf.it}
	\affiliation{INAF/OAS Bologna, via Gobetti 101, I-40129 Bologna, Italy}
	\affiliation{INFN, Sezione di Bologna, Via Berti Pichat 6/2, I-40127 Bologna, Italy}
	
	\author{Mauro Pieroni}\email{mauro.pieroni@cern.ch}
	\affiliation{Theoretical Physics Department, CERN, 1211 Geneva 23, Switzerland}
    
    \date{\today}

    \preprint{CERN-TH-2023-191}
    
    \begin{abstract}
    \noindent
     Gravitational Waves (GW) sourced by second-order primordial curvature fluctuations are among the favored models fitting the recent pulsar timing array (PTA) measurement of a stochastic GW background (SGWB). We study how spectral distortions (SDs) and anisotropies of the cosmic microwave background (CMB) can constrain such scalar fluctuations. Whereas COBE FIRAS data have no sufficient sensitivity to probe the PTA log-normal hypothesis, we show how future PIXIE-like experiments can detect the CMB SDs from the scalar-induced interpretation of the SGWB in PTA data. We finally show how the transformative synergy between PTA data and future CMB SD measurements is important for reconstructing primordial fluctuations at these small scales. 
    \end{abstract}
	
    \pacs{Valid PACS appear here}
    \maketitle

\noindent{\bf Introduction.} 
One of the goals of modern cosmology is the characterization of the primordial power spectrum of the curvature perturbations $\mathcal{P}(k)$ generated during inflation.
The long wavelength quantum fluctuations amplified during inflation classicalize and re-enter the Hubble radius in the radiation and matter eras, providing the initial seeds for the gravitational instability in the large scale structure of the Universe. 

The most stringent constraints on $\mathcal{P}(k)$ arise from Cosmic Microwave Background (CMB) anisotropies measurements, revealing a near scale invariant, slightly red-tilted, spectrum on very large scales within the range $[0.001,\,0.1] \,{\rm Mpc} ^{-1}$. 
The Planck DR3 data constrain the amplitude $A_s$ of $\mathcal{P}(k)$ at $k = 0.05 \,{\rm Mpc}^{-1}$ and its spectral index to $\ln\left(10^{10}\,A_s\right)=3.044 \pm 0.014$ and $n_s=0.9649 \pm 0.0042$ at 68\% CL, respectively~\cite{Planck:2018jri}. 
Galaxy surveys can extend these constraints to $ {\cal O} (1) \,{\rm Mpc} ^{-1}$, but smaller scales remain largely unconstrained.

Recent observations of a Stochastic Gravitational Wave Background (SGWB) at nHz frequencies by Pulsar Timing Arrays (PTA)~\cite{NANOGrav:2023gor,Reardon:2023gzh,EPTA:2023fyk,Xu:2023wog} have sparked a significant interest in $\mathcal{P}(k)$ at much smaller scales, since scalar fluctuations can generate such a SGWB at second order in perturbation theory~\cite{Matarrese:1997ay,Ananda:2006af,Baumann:2007zm}  at scales $[10^7,\,10^9] \,{\rm Mpc} ^{-1}$.  
Recent studies suggest that such a Scalar Induced Gravitational Wave Background (SIGWB) could provide a viable explanation for the PTA detection and might be favored over many other candidates from a Bayesian perspective~\cite{NANOGrav:2023hvm, EPTA:2023xxk} (see however \cite{NANOGrav:2023hvm, EPTA:2023xxk,Franciolini:2023pbf, Figueroa:2023zhu, Ellis:2023oxs} for a discussion of potential PBH overproduction associated to this SIGWB explanation and ~\cite{Franciolini:2023pbf, Figueroa:2023zhu, Ellis:2023oxs, Liu:2023ymk, Wang:2023ost, Cai:2023dls} for alternative analyses). 
If substantiated, PTA measurements could provide valuable insights into the later stages of inflation, with profound implications for theoretical models.
	
\begin{figure*}
	\centering
 \includegraphics[width=.85\columnwidth]{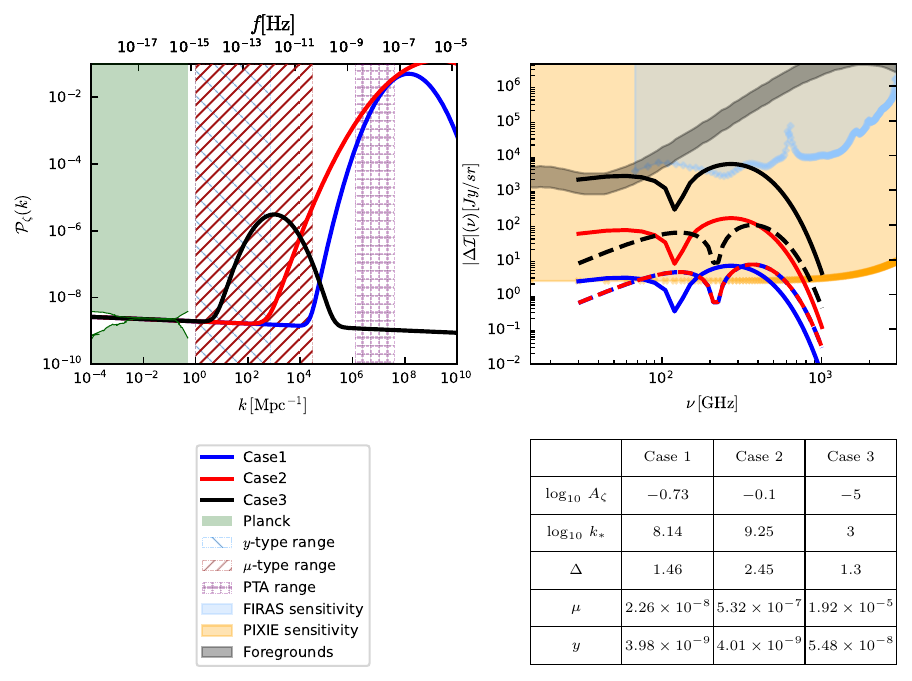}
	\caption{Left: illustrative primordial spectra together with constraints from Planck and sensitivity ranges of $y$ and $\mu$-type distortions and PTA.  Right: corresponding $\mu$ (solid lines) and $y$-type (dashed lines) SD signals as functions of the frequency $\nu$ of the CMB black-body spectrum, compared to the sensitivity of FIRAS and PIXIE as well as to the total contribution of foregrounds. 
    $\mu$ and $y$ distortions are computed by adopting the window functions of PIXIE.  
    In the Table, we report the fiducial parameters for each case. }
	\label{fig:intro}
\end{figure*}
	
It is therefore of utmost importance to devise further tests of this hypothesis, and, as is often the case in cosmology, the synergy with other probes of $\mathcal{P}(k)$ becomes invaluable.
In this context, it is noteworthy that the dissipation into acoustic waves of curvature perturbations responsible for the SIGWB inevitably induces spectral distortions (SDs) of the CMB spectrum by mixing photons with different temperatures \cite{Hu:1994bz,Chluba:2012we,Clesse:2014pna,Kohri:2014lza,Chluba:2015bqa,Byrnes:2018txb,Unal:2020mts,Schoneberg:2020nyg}. 
The scales tested by SDs lie within the range $\sim[1,\,10^4]\,{\rm Mpc}^{-1}$, effectively bridging the observational gap between CMB anisotropies and PTA, thereby extending our leverage in constraining $\mathcal{P}(k)$, as shown in Fig.~\ref{fig:intro}.  
Furthermore, since alternative interpretations of PTA signals, such as those attributed to SMBHs, phase transitions, or cosmic strings do not predict significant SDs, CMB anisotropies and SDs offer an opportunity to falsify the SIGWB hypothesis underlying the PTA detection.

\medskip\noindent
{\bf Motivation and data.} 
In this study, we explore the implications of the SIGWB interpretation of recent PTA data~\cite{NANOGrav:2023gor,Reardon:2023gzh,EPTA:2023fyk,Xu:2023wog} on the anisotropies and SDs of the CMB. We use Planck data for CMB anisotropies, and, for SDs, we consider both COBE FIRAS constraints~\cite{Fixsen:1996nj} and what can be achieved from future spectrometers, like PIXIE~\cite{Kogut:2011xw}.
We use NANOGrav 15 publicly available data~\cite{NANOGrav:2023gor} for the recent PTA measurement of SGWB~\cite{NANOGrav:2023gor,Reardon:2023gzh,EPTA:2023fyk,Xu:2023wog}.

Whereas the complementarity between CMB SDs and PTA data was previously discussed~\cite{Byrnes:2018txb,Unal:2020mts,Schoneberg:2020nyg,Madge:2023cak}, our work is the first one where the synergy between the recent PTA detection of an SGWB~\cite{NANOGrav:2023gor,Reardon:2023gzh,EPTA:2023fyk,Xu:2023wog} and CMB SDs is studied quantitatively in a fully consistent Bayesian way by taking into account marginalization over foreground contamination and reionization effects. Our analysis also differs from previous approaches in technical details. 

Instead of relying on the commonly employed window function approximation for calculating $\mu$ and $y$-type SDs~\cite{Pajer:2012vz,Chluba:2013dna}, we rigorously compute them using the Green's function method proposed in~\cite{Chluba:2013vsa} and implemented in the extension of the \texttt{CLASS} code~\cite{Blas:2011rf} introduced in~\cite{Lucca:2019rxf,Fu:2020wkq}.

 \medskip\noindent
{\bf The analysis.}  
In our analysis, we adopt a straightforward yet widely used lognormal (LN) parametrization for a peak in $\mathcal{P}(k)$~\cite{Pi:2020otn} which encapsulates its critical attributes, including amplitude, location, and width:
\begin{align}\label{eq:lognormal}
	\mathcal{P}_{\rm LN }(k)  =&\frac{A_\zeta}{\sqrt{2\pi}\Delta}\exp\left[-\frac{\ln^2 k/k_*}{2\Delta^2}\right].
\end{align}
While this parametrization may not capture the details of specific inflationary models across all scales, it enables us to draw generic conclusions and to establish connections with recent PTA analyses~\cite{Dandoy:2023jot,NANOGrav:2023hvm,Reardon:2023gzh,EPTA:2023xxk,Figueroa:2023zhu,Franciolini:2023pbf}, which also employed the parametrization in Eq.~\eqref{eq:lognormal}.  
We will comment on possible extensions in the Conclusions. 
We add this LN peak to the $\Lambda$CDM power-law spectrum of curvature perturbation parametrized by $(A_s, n_s)$.

Within the LN parametrization, we adopt 3 fiducials, whose parameters are reported in Fig.~\ref{fig:intro} together with their $\mu$ and $y$ SD signals. 
We chose them as follows:
\begin{enumerate}[(i)]
    \item {\em Case 1} is the maximum likelihood sample from the MCMC chains of the SIGWB analysis~\cite{Footnote1} of the NG15 collaboration~\cite{NANOGrav:2023hvm}, which we take as a representative PTA dataset. It produces an SD signal that is indistinguishable from a near scale invariant spectrum.    
     \item {\em Case 2} lies well within the 68\% CL contours of NG15 and produces an SD signal that is within the reach of the planned sensitivity of PIXIE. 
    \item {\em Case 3} peaks at the center of the window function of both FIRAS and PIXIE, and produces an SD signal that will be detected by PIXIE at high statistical significance, but weak enough to be allowed by FIRAS. Being outside their sensitivity range, the associated GW signal cannot explain the recent PTA detection, which would be ascribed to either SMBHs or other models. This injection point serves the purpose of depicting the potential of PIXIE alone in constraining $\mathcal{P}(k)$.
\end{enumerate}
We will discuss potential issues associated with these fiducial models in the Conclusions.

\begin{figure*}
	\centering
    \includegraphics[width=0.32\columnwidth]{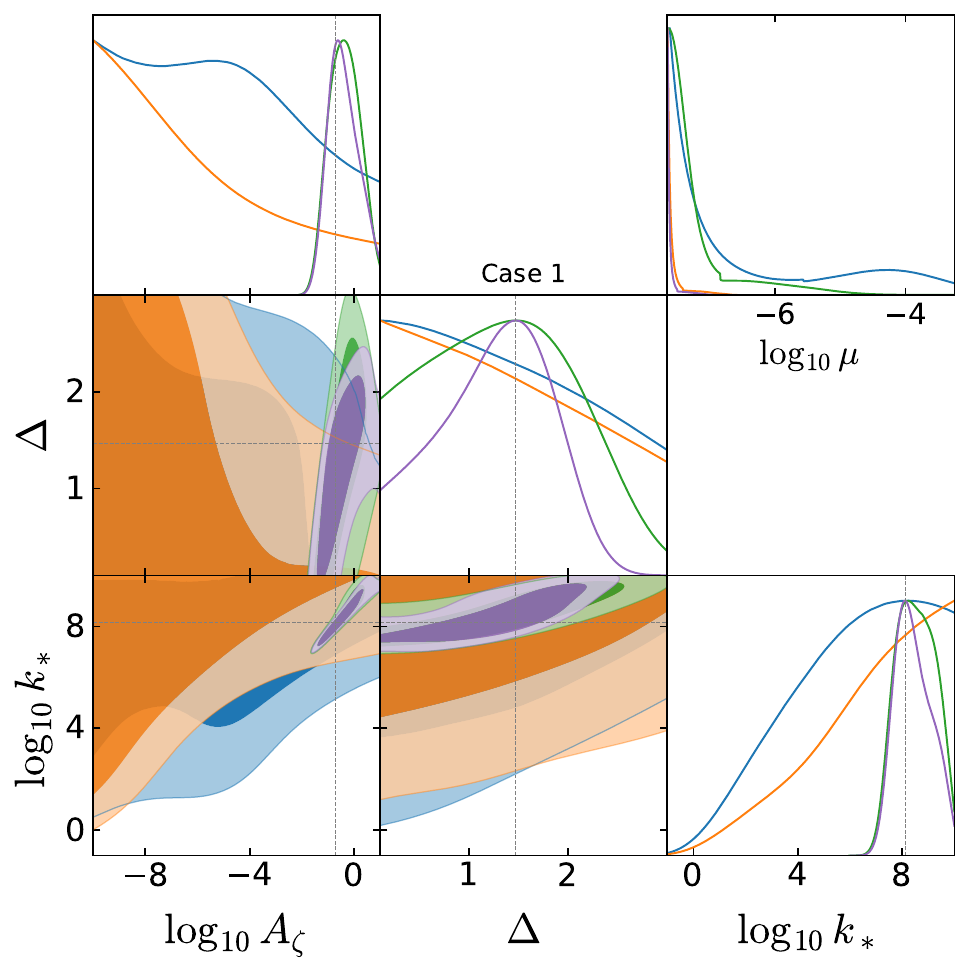}
    \includegraphics[width=0.32\columnwidth]{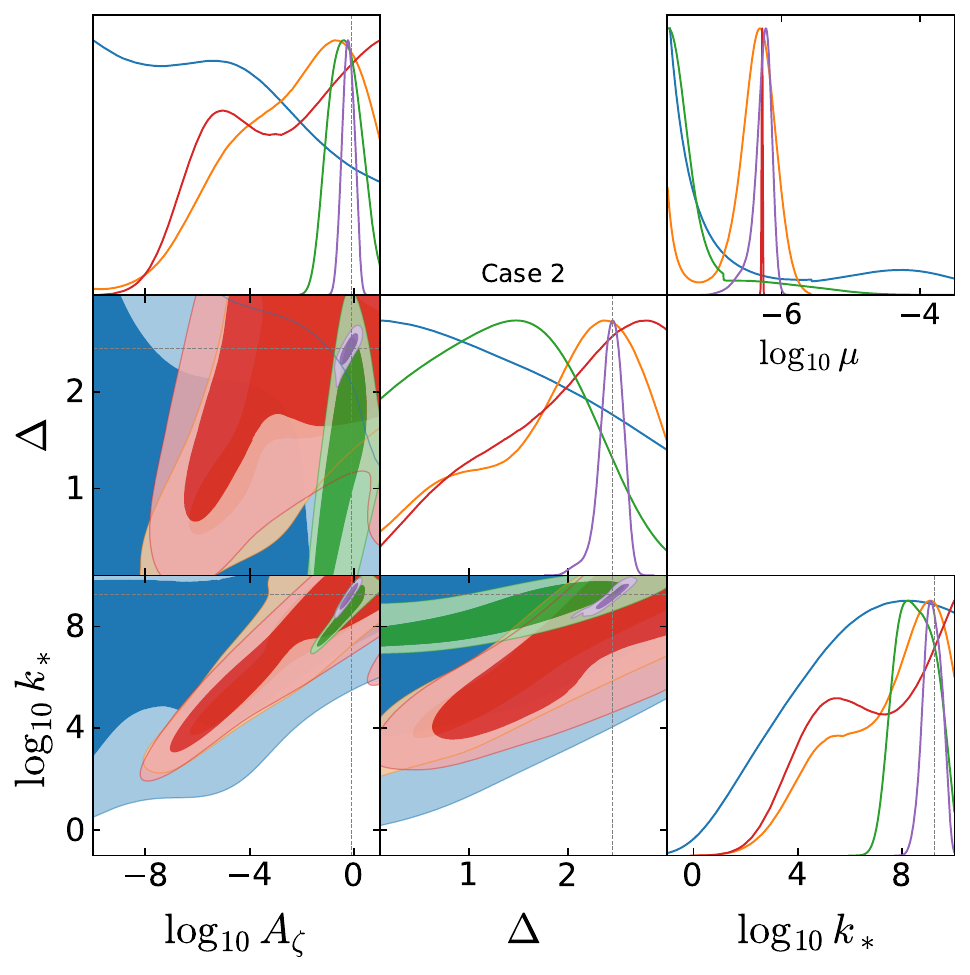}
    \includegraphics[width=0.32\columnwidth]{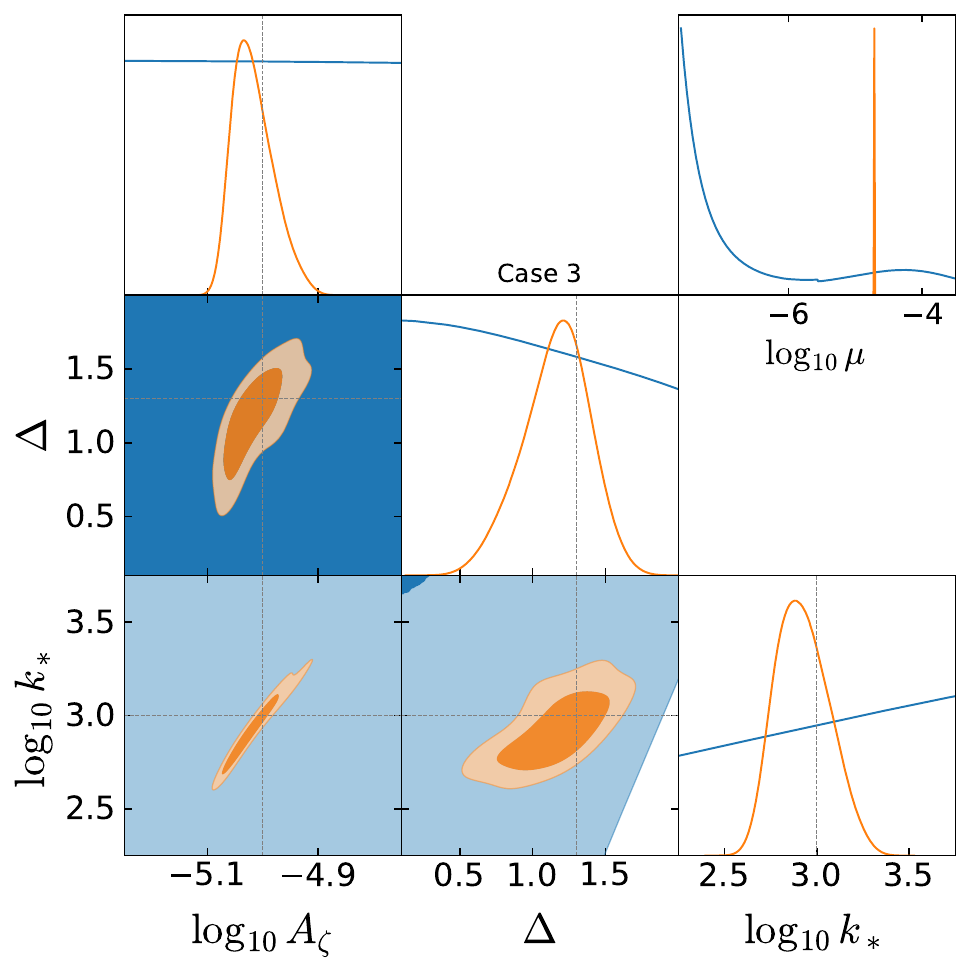}
	\includegraphics[width=\columnwidth]{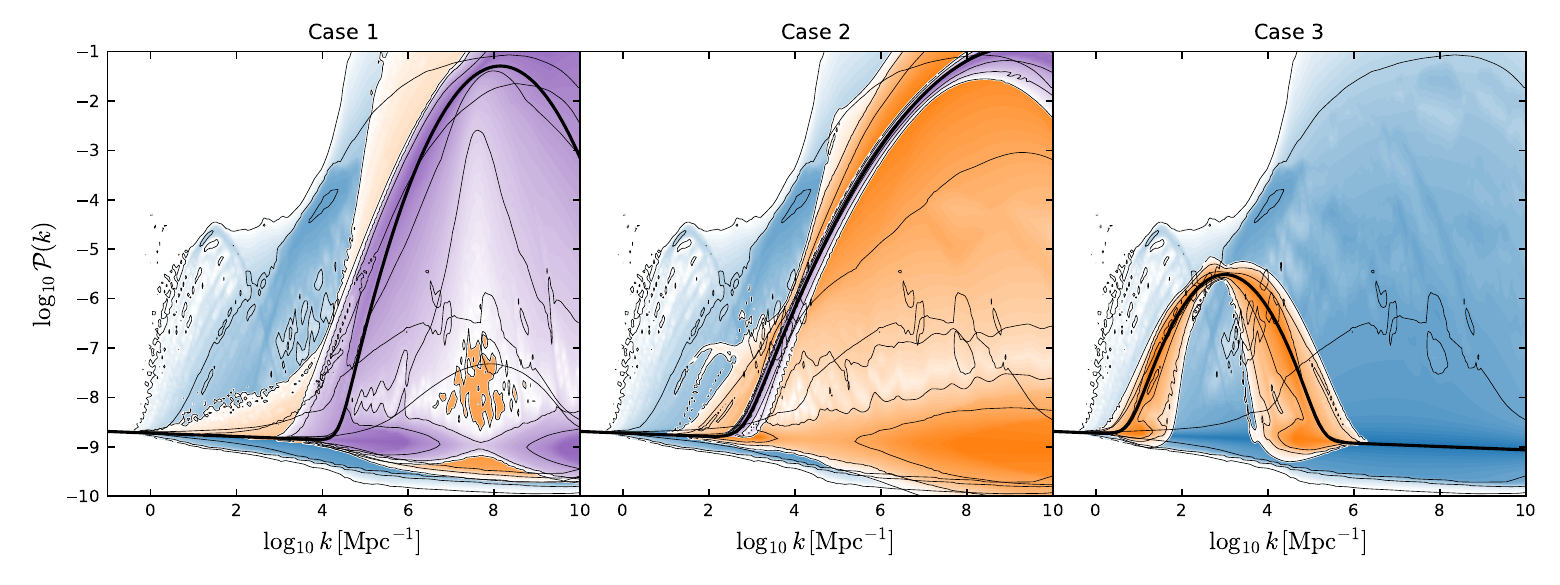}	
    \includegraphics[width=\columnwidth]{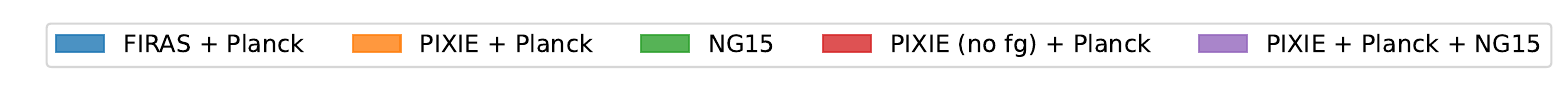}
	\caption{Top: 1D and 2D posterior distributions of the LN parameters, and 1D distribution for the derived parameter $\mu$. The dashed black lines in the triangle plots represent the power spectrum fiducial parameters assumed in each case. Bottom: predictive posterior distributions for $\mathcal{P}_{\rm LN}(k)$. The color shades range from $0$ (darker shades) to $3\,\sigma$ (lighter shades). Constraints from PIXIE are clearly seen to be tighter than those from FIRAS due to the better sensitivity, and the $\mu$ distortion signal is detected in {\em Case} 2 and 3, but not in {\em Case} 1. }
	\label{fig:triangles}
\end{figure*}

We will assume the signal described by each fiducial in turn, and forecast how large the deviations from it can be constrained. The methodology of our MCMC analyses and forecasts is described in detail in the Supplemental Material. 
 There, we also report the priors on the parameters and explain our method of marginalization over nuisance and foreground parameters in the case of SDs. 
 In addition to presenting 1d and 2d constraints on the model parameters, we also plot the predictive posterior distribution of $\mathcal{P}(k)$, allowing a straightforward visualization of the constraints on $\mathcal{P}(k)$ itself, which cannot be grasped from the triangle plots alone. 
 
\medskip\noindent
{\bf Constraints from FIRAS and Planck.} 
For all the three cases studied, FIRAS data are not very informative, as can be seen from the blue contours in the triangle plots in Fig.~\ref{fig:triangles}, and in combination with Planck can only exclude some regions of the parameter space - in particular, Planck provides a lower limit for $k_*$. 
Because our prior on $k_*$ allows the peak in $\mathcal{P}(k)$ to also fall completely outside the window function of FIRAS, the amplitude $A_\zeta$ is unconstrained after marginalizing over the other parameters. 
However, we can clearly see a {\em scale dependent} bound on $A_\zeta$ in the 2D plane $\log_{10}\,A_\zeta$ vs $\log_{10}\,k_*$. 
The posterior distribution of $\mathcal{P}(k)$ depicts tight constraints at the largest scales tested by Planck~\cite{Planck:2018jri}. 
Moving to intermediate scales, within the window function of FIRAS, the upper limit on the amplitude is roughly $A_\zeta\lesssim 10^{-4}$. 
At smaller scales, to which FIRAS is not sensitive, $\mathcal{P}(k)$ becomes unbounded. 
Our results are consistent with those of previous analyses~obtained with simpler approximations~\cite{Byrnes:2018txb,Unal:2020mts}.

\medskip\noindent
{\bf Forecasts for PIXIE plus Planck.} 
Our PIXIE forecasts in combination with Planck are shown in Fig.~\ref{fig:triangles}, on top of the constraints from FIRAS. 
PIXIE and Planck cannot detect the signal for {\em Case 1}, mainly due to the smallness of the predicted $\mu$ parameter. 
While, due to the much better sensitivity, the constraints on the amplitude do improve, the bump in the power spectrum is located at such small scales and does not contribute to the SD signal. Furthermore, we highlight that the scale at which large deviations from near scale invariance are allowed is pushed to $>10^{4}$ Mpc$^{-1}$.
{\em Case 2} is perhaps the most interesting one (see also the Supplemental Material for more details).
We see that, despite the predicted $\mu$ being well within the reach of the nominal PIXIE sensitivity, 
i.e., $\mu_{\rm PIXIE}\lesssim 2\times10^{-8}$ as quoted in~\cite{Kogut:2011xw}, the model parameters are not well constrained and, although the reconstructed spectrum does show a hint of a deviation from near scale invariance, the bump is not detected and $\mathcal{P}(k)$ is still consistent with a near scale invariant one.  
$\mu$ and $y$ distortions are indeed consistent with the minimal signal predicted by $\Lambda$CDM, although the posterior distribution of $\mu$ does show a peak around the fiducial value of $\mu=5.32\times10^{-7}$. 
This agrees with previous studies (see e.g.~\cite{Fu:2020wkq}) that indicate that the PIXIE nominal sensitivity is degraded by almost an order of magnitude when including foreground contamination~\cite{SathyanarayanaRao:2015vgh,Desjacques:2015yfa,Mashian:2016bry,Abitbol:2017vwa,Rotti:2020tlo,Zelko:2020ojo}. 
As seen from the red contours in Fig.~\ref{fig:triangles}, removing foregrounds would allow a significant detection of $\mu = \left(\,5.27\pm 0.09\,\right)\times 10^{-7}$ at 68\% CL, although 
constraints on $\mathcal{P}(k)$ are nearly unchanged.
We note that either way, the near scale invariant spectrum will be ruled out, as it cannot explain such a large $\mu$ signal. Another way to see it is the lower bound we get on the LN amplitude $\log_{10}\,A_\zeta > -7.2$ at 95\% CL, not consistent with a near scale invariant spectrum.  We thus learn two valuable lessons. On one hand, foregrounds can degrade the sensitivity of SD experiments to the primordial power spectrum. 
On the other hand, even after foreground removal and despite ruling out a near scale invariant spectrum, if the SD signal is not strong enough so that both $\mu$ and $y$ distortions are detected, it becomes complicated to detect with high confidence the LN model parameters
\eqref{eq:lognormal}, due to the parameter degeneracies. Indeed, as suggested by our analytical approximations of $\mu$ and $y$ in Eq.~(9) of the Supplemental Material, different combinations of $(A_\zeta,\,k_*,\,\Delta)$ can lead to the same $(\mu,\,y)$. This is also the reason why we do not observe a decisive detection of an LN peak in the reconstructed $\mathcal{P}(k)$. Since $k_*$ is not detected, there is no uniquely defined locus of curves that all have a peak at the same region, and a large fraction of the $(\mathcal{P}(k),\,k)$ is allowed.\footnote{We refer the reader to Fig.~3 of our Supplemental Material for more details on this point.}
While estimating $\mu$ and $y$ distortions can quickly indicate if there is hope to detect a signal with future experiments (see e.g.~\cite{Byrnes:2018txb,Unal:2020mts}), 
only with our MCMC pipeline, which also takes into account the impact of foregrounds, can we rigorously and robustly forecast constraints on the model parameters controlling $\mathcal{P}(k)$. 
This is one of the main results of this work.
Finally, the SD signal from {\em Case 3} is so loud that PIXIE can detect  $\mu=\left(\,1.92\pm 0.02\,\right)\times 10^{-5}$
and $y=\left(\,4.9^{+1.7}_{-3.1}\,\right)\times 10^{-8}$ at 68\%CL. 
As shown in the corresponding triangle plot, while some degeneracies between the model parameters still show up, the model parameters can be very well constrained and a near scale invariant $\mathcal{P}(k)$ is excluded with extremely high confidence in the range $k\in[10^2,\,10^4]\,{\rm Mpc}^{-1}$. 

\medskip\noindent
{\bf Synergy with PTA.} 
Let us now discuss the synergy between future SD experiments and PTA data. 
To do that, we take the multi-dimensional posterior distribution on the LN parameters $(\log_{10}\,A_\zeta,\,\Delta,\,\log_{10} k_*)$ from the public chains of the NG15 analysis of the SIGWB and use it as input for our forecast for PIXIE. 
We discussed how {\rm Case 1} cannot be detected by PIXIE, as its SD signal is too weak. 
However, the parameter space allowed by NG15 alone still contains some regions producing a large $\mu$-type distortion, which will be ruled out by PIXIE, as seen by comparing the green and purple contours in the triangle plot for {\rm Case 1}.
The improvement in the constraints is more 
visible for {\em Case 2}. 
Despite consistently marginalizing over foregrounds and temperature shifts, the fiducial parameters can be recovered and constrained for this case. This can be physically understood by the fact that $\mu$ and $\Omega_{\rm GW}$ depend on convolutions and squares of $\mathcal{P}(k)$, respectively, and are thus sensitive to different combinations of the model parameters, which can be appreciated by comparing the orange/red and green contours in the central triangle plot in Fig.~\ref{fig:triangles}, where degeneracies run along different directions. 
Therefore the synergy between SDs and PTA could help break such parameter degeneracies and could be thus extremely helpful to probe large deviations in $\mathcal{P}(k)$ from a power-law shape beyond the detection of $\mu$.

\begin{figure*}
	\centering
	\includegraphics[width=0.90\columnwidth]{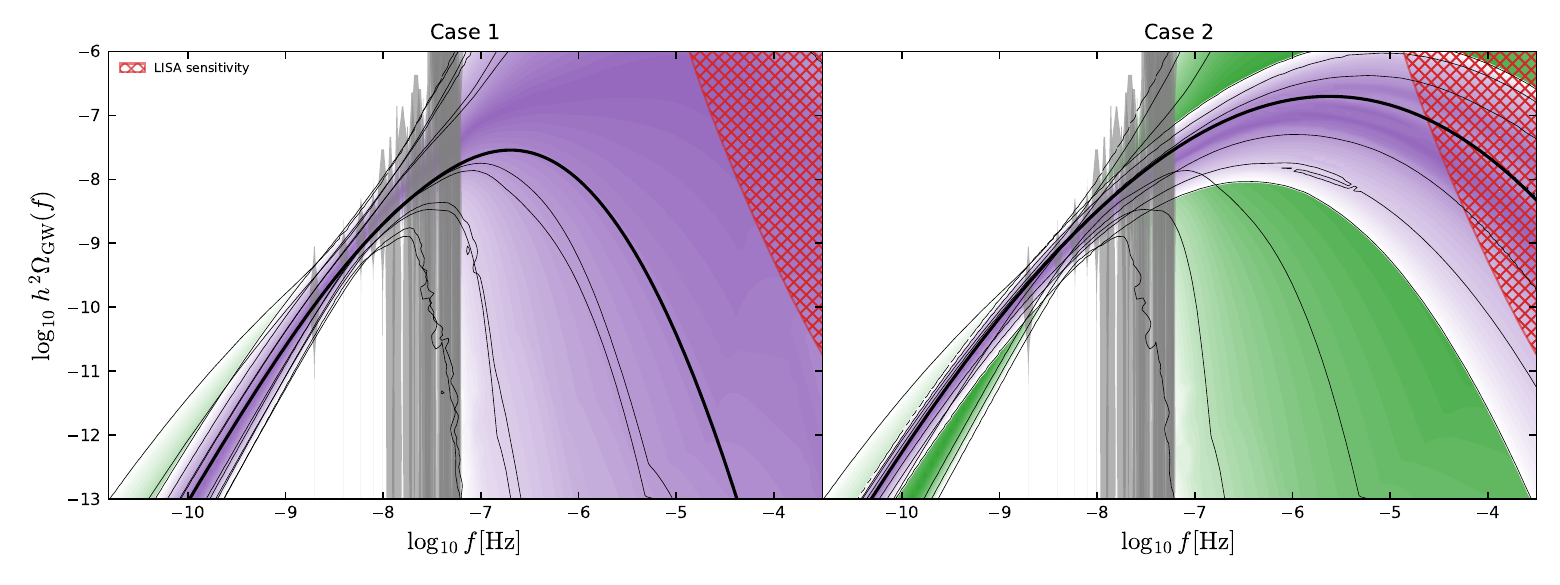}
	\caption{Derived constraints on the spectral shape of the SGWB from the NG15 analysis (green) as well as the forecasts for the PIXIE + Planck + NG15 in {\em Case 2} (purple). We also plot the NG15 violin plots and the LISA power-law sensitivity curve at higher frequencies. The color shades of the posterior probability of $\Omega_{\rm GW}$ range from $0$ (darker shades) to $3\,\sigma$ (lighter shades).}
	\label{fig:Omega}
\end{figure*}

\medskip\noindent
{\bf Conclusions.} 
Motivated by the recent detection of an SGWB by PTA experiments, and its possible scalar-induced nature~\cite{NANOGrav:2023gor,Reardon:2023gzh,EPTA:2023fyk,Xu:2023wog,NANOGrav:2023hvm, EPTA:2023xxk}, we investigated how future measurements of CMB SDs could probe the dissipation into acoustic waves of these scalar perturbations producing the SGWB. 
We found that the parameter space consistent with PTA observations spans several orders of magnitude in $\mu$-type and $y$-type SD signals, with both small and large amplitudes. 
While COBE FIRAS data contribute minimally to the constraints within this range, future experiments such as PIXIE hold great promise for probing this uncharted territory, especially when combined with PTA data. 
Furthermore, we showed that an experiment like PIXIE will be able to bridge between the scales tested by CMB anisotropies and PTA, probing corners of the parameter space not accessible to either of them and thus a unique target of future SD experiments. 
Note also that our results are somewhat conservative since more sensitive concepts than PIXIE have been proposed as future space experiments for CMB SDs. An experiment such as Super-Pixie would achieve far better errors $\sigma(y)\simeq1.6\times10^{-9}$ and $\sigma(\mu)\simeq7.7\times10^{-9}$ even after foregrounds marginalization, and the sensitivity of ESA Voyage 2050 concept would be up to 5 times better than Super-PIXIE~\cite{Chluba:2019nxa}. These advancements would enable more stringent constraints than those presented in this paper, allowing SDs tests of models such as our {\em Case 1}. 

We also identify promising avenues that go beyond the results presented in this paper:
\begin{enumerate}[(i)]
    \item {\em Beyond the LN template.} Our choice for the LN template as workhorse for our analysis has been dictated both for simplicity and because it has been chosen by the NANOGrav collaboration.  Features beyond the LN template have been proposed, such as different infrared or ultraviolet slopes, large dips characteristic of ultra-slow-roll dynamics~\cite{Balaji:2022zur}, steep growth of the infrared tail beyond $k^4$, indicative of multifield scenarios~\cite{Byrnes:2018txb,Fumagalli:2020adf,Palma:2020ejf,Braglia:2020eai,Fumagalli:2020nvq,Braglia:2020taf,Braglia:2022phb,Cole:2022xqc}, or large oscillations due to sharp features~\cite{Fumagalli:2020adf,Palma:2020ejf,Braglia:2020eai,Fumagalli:2020nvq,Braglia:2020taf}.  Repeating our analyses with more realistic templates or direct numerical integration of inflationary equations can bridge the gap between data and fundamental theoretical parameters.
    \item {\em PBHs.} A large peak in $\mathcal{P}(k)$ may lead to the formation of PBHs. A rough estimate of the threshold for (over)production of PBHs is $A_\zeta\sim\mathcal{O}(0.01-0.1)$. Our posteriors include amplitudes much larger, raising questions about the overproduction of light PBHs. There are, however, several well known caveats that affect the quick estimates above, which again rose to prominence after the PTA detection. The PBH abundance is strongly affected by the shape of $\mathcal{P}(k)$~\cite{Gow:2020cou}, the equation of state (EOS) of the Universe at PBH formation~\cite{Byrnes:2018clq} and, notably, non-Gaussianities~\cite{Ferrante:2022mui,Gow:2022jfb,Franciolini:2023pbf,Inomata:2023drn}. The SIGWB would also be modified upon including changes in the equation of state of the Universe~\cite{Franciolini:2023wjm,Balaji:2023ehk,Harigaya:2023pmw} and non-Gaussianities in the calculation of the SGWB~\cite{Unal:2018yaa,Adshead:2021hnm,Ragavendra:2021qdu,Garcia-Saenz:2022tzu,Li:2023xtl,Cai:2018dig}. In particular, it is worth mentioning here that large non-Gaussianities could simultaneously suppress PBH formation, alleviating tensions about their overproduction~\cite{Franciolini:2023pbf}, and enhance the SD signal~\cite{DeLuca:2021hcf,Sharma:2024img}. In addition to these model dependencies, other sources of uncertainties such as the PBH formation criteria~\cite{Gow:2020bzo,Inomata:2023zup} and the accurate size of the physical horizon at formation~\cite{DeLuca:2023tun} may also affect our conclusions regarding the overproduction of PBHs. Lastly, we also mention that PBHs would impact SDs also through their accretion~\cite{Ali-Haimoud:2016mbv}. A state-of-the-art discussion of the resulting constraints that derive from PBH formation is beyond the scope of this work.
    \item {\em Other CMB SD imprints.} Finally, although we have only considered the total intensity spectrum of the CMB, other SD observables such as  $T-\mu$ correlation~\cite{Pajer:2012vz,Ozsoy:2021qrg,Bianchini:2022dqh} and SD anisotropies~\cite{Zegeye:2021yml} have been proved to provide valuable constraints on the shape of $\mathcal{P}(k)$. Furthermore, tensor perturbations also directly source SDs~\cite{Ota:2014hha,Chluba:2014qia}, although their signal is much weaker compared to the one explored in this paper. All such insights add to the prospects of testing the primordial power spectrum with future SD experiments.    
\end{enumerate}

Our results also demonstrate that measuring an SD signal will advance our understanding of primordial physics and also hold significant implications for interpreting data from future space-based GW observatories, such as LISA~\cite{LISA:2017pwj} or BBO~\cite{Crowder:2005nr}-Decigo~\cite{Kawamura:2006up}, as displayed in Fig.~\ref{fig:Omega}. 
When analyzing current PTA data in terms of an LN peak, the predicted $\Omega_{\rm GW}$ from the posterior distribution of the model parameters intersects the sensitivity of future GW experiments. 
Adding information from SD anchors a SIGWB at large scales, so that its shape within the LISA sensitivity band is theoretically predicted. 
This theoretical prior can optimize the search of an SGWB at ${\rm mHz}$ frequencies, a key scientific target of LISA~\cite{ Bartolo:2016ami, Caprini:2019egz, LISACosmologyWorkingGroup:2022jok}. We believe that such profound physical implications provide strong support for the science case of a future space-based CMB spectrometer. We intend to explore these avenues further in future publications, building upon the foundation laid out in this study.

\medskip
\noindent{\em Softwares.} We made use of the publicly available softwares \texttt{CLASS}~\cite{Blas:2011rf}, \texttt{MontePython}~\cite{Brinckmann:2018cvx}, \texttt{GetDist}~\cite{Lewis:2019xzd} and \texttt{fgivenx}~\cite{fgivenx}.

\medskip
\noindent{\em Note added.} While this project was near to completion, a related paper~\cite{Cyr:2023pgw}, also studying the capability of SDs to constrain $\mathcal{P}(k)$ and its relation to the recent PTA detection,  appeared on the arXiv.
\medskip

\begin{acknowledgments} 
\noindent{\bf Acknowledgments.} 
We thank Chris Byrnes, Jens Chluba, and Matteo Lucca for interesting discussions at various stages of this project, and Chris Byrnes and Valerie Domcke for comments on a draft of the paper. We acknowledge the use of computational resources from the parallel computing cluster of the Open Physics Hub (\url{https://site.unibo.it/openphysicshub/en}) at the Physics and Astronomy Department in Bologna and from the INAF/OAS Bologna cluster. MT acknowledges the funding by the European Union -  NextGenerationEU, in the framework of the HPC project – “National Centre for HPC, Big Data and Quantum Computing” (PNRR - M4C2 - I1.4 - CN00000013 – CUP J33C22001170001)). FF acknowledges financial support from the INFN InDark initiative and from the COSMOS network ({\tt www.cosmosnet.it}) through the ASI (Italian Space Agency) Grants 2016-24-H.0 and 2016-24-H.1-2018, as well as 2020-9-HH.0 (participation in LiteBIRD phase A).  MB thanks the Theoretical Physics Department of CERN and INAF/OAS Bologna, where part of this work was carried out, for hospitality. MP acknowledges the hospitality of Imperial College London, which provided office space during some
parts of this project.
\end{acknowledgments}

\newpage
\begin{widetext}

\appendix

\section{Supplemental material}

\noindent{\bf CMB spectral distortions.}
Here we briefly review the calculation of the SD signal~\cite{MTthesis}, focusing on the imprints of the shape of the PPS on $\mu$ and $y$-type distortions. 
The SD of the CMB $\Delta I$ can be decomposed as follows:
\begin{equation}   \label{eq:delta-I} 
\Delta I(x) = \Delta I_{\rm SZ}(x) + \Delta I_T(x) + \Delta I_{\rm fg}(x) + \Delta I_{\rm exotic} + \Delta I_{\rm \Lambda CDM},
\end{equation}
where $x$ is the dimensionless frequency $x = h\nu/(k_B T_0)$, and the different terms represent contributions from late-time reionization $\Delta I_{\rm SZ}$, temperature shift  $\Delta I_{\rm T}$, foregrounds $\Delta I_{\rm fg}$, possible exotic physics $\Delta I_{\rm exotic}$ and the minimal signal produced by $\Lambda$CDM $\Delta I_{\rm \Lambda CDM}$.
Let us briefly comment on each of these term.

${\bf \Delta I_{\rm SZ}(x)}$ is the late-time Sunyaev-Zeldovich contribution from energetic reionized electrons that scatter CMB photons creating SDs. 
This term can be approximated as in~\cite{Schoneberg:2020nyg}:
\begin{equation}
    \Delta I_{\rm SZ}(x) = \mathcal{N}x^3 y_{\rm reio} Y(x).
\end{equation}
 ${\bf\Delta I_{\rm T}}$ arises from the difference between the true CMB temperature $T$ and its reference value $T_0$:
\begin{equation} \label{eq:sd_deltaT}   
 \Delta I_T(x) = \mathcal{N}x^3\left[\Delta_T(1 + \Delta_T)G(x) + (\Delta_T^2/2)Y(x) \right] ,
\end{equation}
where $\mathcal{N} = 2(k_B T)^3/(hc)^2$, $G(x) = xe^x / (e^x -1 )^2$, $Y(x) = G(x) [x (e^x + 1)/(e^x -1) - 4 ]$ and $\Delta_T = (T - T_0)/T_0$.

 ${\bf\Delta I_{\rm fg}}$ encodes foreground contributions from the Galactic thermal dust, the Cosmic Infrared Background (CIB),  synchrotron, free-free, spinning dust, and integrated CO emissions. 
A detailed description of this term and its implementation in \texttt{MontePython} can be found in~\cite{Abitbol:2017vwa, Schoneberg:2020nyg}.

${\bf \Delta I_{\rm exotic}}$ contains the cosmological signal from exotic energy injections at high redshifts. As we are not interested in such Physics in this paper, we set it to $\Delta I_{\rm exotic} = 0$.

Finally, ${\bf  \Delta I_{\Lambda {\rm CDM} } }$ is the sum of two main contributions, i.e. adiabatic cooling of electrons and baryons and dissipation of acoustic waves. Since the former effect is not related to the PPS, we describe only the latter. The presence of primordial density fluctuations generated during inflation causes some regions of the primordial plasma to be hotter and denser than others. 
The various regions of the plasma at different temperatures are all represented by a black body spectrum.
Photons diffuse from overdense to underdense regions and vice versa, generating an isotropization of the photon phase space distribution at small scales~\cite{Silk:1967kq} and CMB SDs~\cite{Sunyaev:1970plh, Daly:1991uob, 10.1093/mnras/248.1.52, Hu:1994bz}.
The rate of the energy injected in the plasma due to the Silk damping is ~\cite{Chluba:2013dna}
\begin{equation}\label{eq:sd-enery-rate-pps}
    \f{\dd Q}{\dd t} \approx 4 A^2 \rho_\gamma \int \dd k \, k \, \mathcal{P}(k) \, (\partial_t k^{-2}_D ) \, e^{-2(k/k_D)^2},
\end{equation}
where $k_D$ is the Silk-damping scale, $\rho_\gamma$ the photon density and $A \approx (1 + (4/15)\rho_\nu / \rho_r)^{-1}$. 
We therefore see that the energy rate is proportional to the PPS, which sets the initial condition for the photon perturbations in the Early Universe. 
In general, the total SD signal can be decomposed as the sum of three different types of SDs, namely $\mu$, $y$, and $g$, and some residual term $R(x)$ that represents all the SDs that are neither $\mu$, $y$ or $g$ distortions.  
Each contribution to the SD signal is characterized by an amplitude and a shape. 
The $y$ and $g$ shapes are denoted by $Y(x)$ and $G(x)$ respectively, while the $\mu$ shape is $M(x) = G(x) [0.4561 - 1 /x]$.
Therefore, $\Delta I_{\Lambda \rm CDM}$ can be written as 
\begin{equation}
    \Delta I_{\Lambda {\rm CDM} } (x) =  g G(x) + \mu M(x) + y Y(x) + R(x) 
\end{equation}
The SD amplitudes $\mu$, $y$, and $g$ are in general computed given the rate of the energy into photons $\dd Q / \dd t$ as:
\begin{equation}\label{eq:sd-ampl}
    a = \int_0^{\infty} \dd z \f{\dd Q / \dd t}{(1+z) H \rho_\gamma}\mathcal{J}_a(z), \qquad \text{for} \qquad a = \mu, y, g,
\end{equation}
where $\mathcal{J}_a(z)$ describes how much of the injected energy creates an $a$-type SD at $z$.
There are different ways to compute the visibility functions $\mathcal{J}_a(z)$.
We compute them with the PCA method implemented in \texttt{CLASS}~\cite{Lucca:2019rxf}, in which $\mathcal{J}_a(z)$ are equal to the branching ratios of the Green's function of the thermalization problem $G_{\rm th}$ and where $R(x)$ is the residual of $G_{\rm th}$ \cite{Chluba:2013vsa}.
In the acoustic waves dissipation case, the injected energy rate is given by Eq.~\eqref{eq:sd-enery-rate-pps} and therefore the SD amplitudes \eqref{eq:sd-ampl} become
\begin{equation}\label{eq:sds-from-pps}
    a =  \int \f{\dd k}{k}\, \mathcal{P}(k) \, W_a (k),  \end{equation}
 where we define the window functions $W_a(k)$ as
 \begin{equation}\label{eq:windows}
    W_a(k) = 4 \, A^2\, k^2\, \int_0^\infty \dd z \, \f{(\partial_t k^{-2}_D) \,e^{-2(k/k_D)^2} }{H\,(1+z)}\, \mathcal{J}_a(z).
\end{equation}
If the visibility functions $\mathcal{J}_a(z)$ are computed with the PCA method, numerical integration is typically required to calculate the window functions.
However, it is possible to approximate analytically $\mathcal{J}_a(z)$ and perform a straightforward integration of Eq.~\eqref{eq:windows}. 
In this case, it turns out that the approximated $\mu$ window function has support $\sim[1,\,10^4]\,{\rm Mpc}^{-1}$ plotted in Fig.~1 in the main text~\cite{Chluba:2013dna}.

To visualize how $\mu$ and $y$ depend on the template parameters (see Eq.~1 in the main text), we can approximate $W_{\mu,y}(k)$ to  top-hat functions $W_{\mu,y}=A_{\mu,\,y}\times\Theta(k-k_{\rm min})\,\Theta(k_{\rm max}-k)$, where $A_{\mu, y}$ is some proportionality constant.
In this case, the expressions for $\mu$ and $y$ \eqref{eq:sds-from-pps} can be analytically integrated
\begin{equation}\label{eq:simple-mu}
    \mu,\,y = \f{A_{\mu,\,y} \, A_\zeta}{2}\, \l. \erf\l(\f{\ln k - \ln k_*}{\sqrt{2}\, \Delta}\r) \r|^{k^{\mu,\,y}_{\rm max}}_{k^{\mu,\,y}_{\rm min}} 
\end{equation}
The analytic approximation captures the dependence of $\mu$ and $y$ on the LN parameters quite well, except when the LN pivot scale is moved closer to the edges of the window function. 
We show this in Fig.~\ref{fig:mu-class-vs-simple}, where, for simplicity, we compare the analytic result (setting $A_\mu = 2.27$ following~\cite{Byrnes:2018txb}) to the one computed with the PCA method implemented in \texttt{CLASS} assuming the window function of COBE FIRAS. 

\begin{figure*}
	\centering
	\includegraphics[width=0.9\columnwidth]{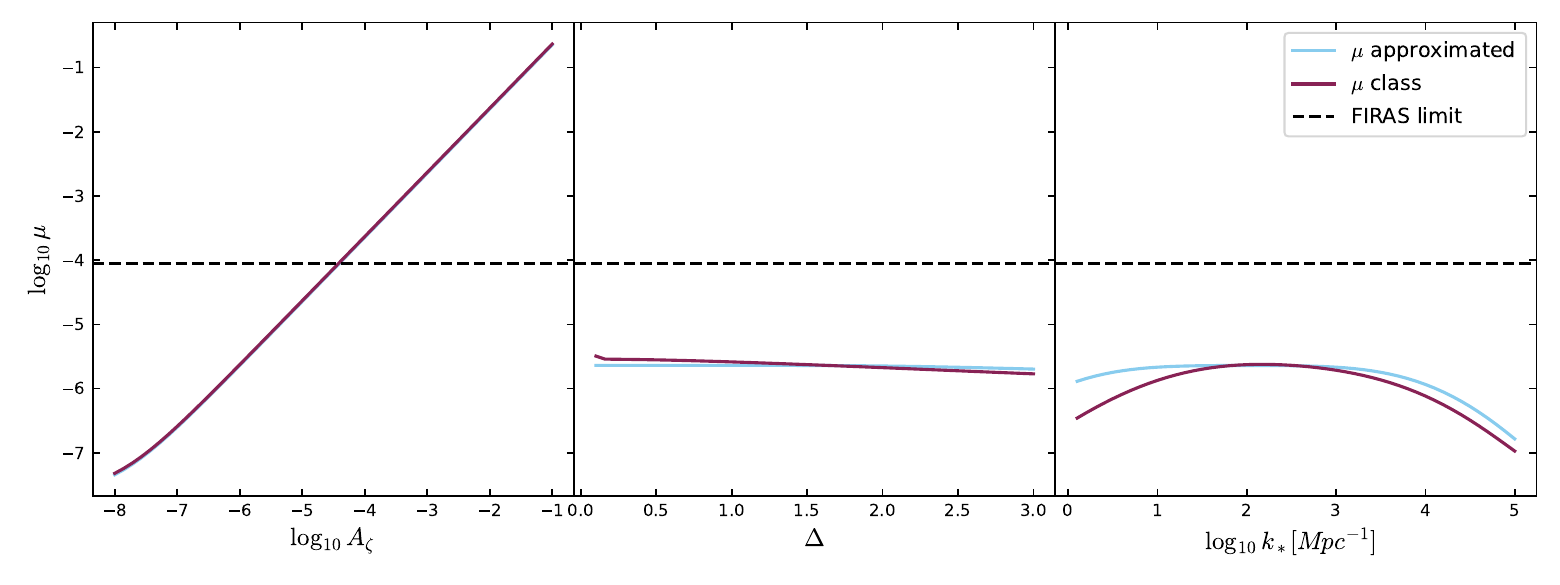}	
	\caption{Comparison between the $\mu$ computed with the PCA method using \texttt{CLASS} and the simple analytic approximation~\eqref{eq:simple-mu}. [Left] we vary $A_\zeta$ while we fix $k_* = \SI{100}{\per\mega\parsec}$, $\Delta = 1.5$. [Center] we vary $\Delta$ and we set $A_\zeta = 10^{-6}$ and $k_* = \SI{100}{\per\mega\parsec}$. [Right] we vary $k_*$ while keeping fixed $A_\zeta = 10^{-6}$ and $\Delta = 1.5$. }
	\label{fig:mu-class-vs-simple}
\end{figure*}

\medskip
\noindent
{\bf CMB SD likelihoods and MCMC methodology.}
In this paper, we place constraints on the PPS through a Bayesian likelihood analysis.
In order to use FIRAS constraints~\cite{Fixsen:1996nj} and to forecast the capability of a future PIXIE-like spectrometer~\cite{Kogut:2011xw}, we use the mock SD likelihood presented in~\cite{Lucca:2019rxf} and implemented in \texttt{MontePython v3.5.0}, which reads as follows
\begin{equation}
\label{eq:SD_likelihood}
    \ln \mathcal{L}_{\rm SD} = -\f{1}{2}\sum_i \l( \f{\Delta I_{\rm obs}(\nu_i) - \Delta I_{\rm pred}(\nu_i)}{\delta I (\nu_i)}\r)^2 \,.
\end{equation}
Here $\Delta I (\nu_i)$ is the detector sensitivity in the $i$-th frequency bin and $\Delta I_{\rm predicted}$ is the theoretical SD signal computed for some given model parameters. In the case of the LN bump in the PPS in  Eq.~1 in the main text, the parameters are $(A_\zeta, k_*, \Delta)$.
For COBE FIRAS, in absence of a publicly available module with real data, we use the mock likelihood implemented in \texttt{MontePython} which is able to reproduce the upper limit on $\mu$ set by the COBE FIRAS~\cite{Fixsen:1996nj}.
$\Delta I_{\rm obs}(\nu_i)$ is the measured (fiducial) SD signal in the $i$-th frequency bin. 
We break the almost perfect degeneracy between $\mu$ and $\Delta_T$ (see Eq.~\eqref{eq:delta-I}) using a Gaussian prior on $\Delta_T$ centered at $\overline{\Delta}_T = 0$, with standard deviation $\sigma_{{\Delta}_T} = 2.2 \times 10^{-4}$ following~\cite{Fu:2020wkq}. 
The Gaussian prior is essential to mitigate the degeneracy, especially in the \emph{Case 2}, where the expected signal $\mu \approx 10^{-7}$ may be suppressed by a value of $\Delta_T$ of the same order. 
Eq.~\eqref{eq:sd_deltaT} shows that the term $\Delta I_T(x)$ includes a term proportional to the $g$-shape, $\Delta_T \cdot G(x)$, so that the overall amplitude multiplying $G(x)$ is the sum of $g$ and $\Delta_T$. 
Typically, the magnitude of $g$ is much smaller, around $10^{-10}$, compared to $\Delta_T$, making $g$-distortions negligible when considering temperature shifts.
For this reason, we follow the default implementation in  \texttt{MontePython} and neglect $g$-distortions in our analysis.
The priors on $y_{\rm reio}$ and the other foreground parameters are also set following~\cite{Schoneberg:2020nyg}.

In order to include constraints on the PPS at $k \lesssim 0.1 \, $/Mpc, we also fold in the constraining power of Planck data. 
Since FIRAS is implemented as a mock likelihood, in order to avoid combining real and simulated data, we use a mock likelihood for Planck reproducing the Planck 2018 $\Lambda$CDM results for cosmological parameters.
The inverse Wishart log-likelihood $\ln \mathcal{L}_{\rm Planck}$ for TT, TE, and EE spectra covers a range of multipoles $\ell \in [2, 2500]$, adopts a modified version of the sensitivity described in \cite{Schoneberg:2020nyg} for the frequency channels 100, 143, and 217 GHz and covers a sky fraction $f_{\rm sky} = 0.65$. 

The results obtained with $\ln \mathcal{L}_{\rm SD}+\ln \mathcal{L}_{\rm Planck}$ by using a Metropolis-Hastings MCMC algorithm implemented in the \texttt{MontePython} sampler and by allowing to vary both the 6 $\Lambda$CDM and 3 LN parameters are shown in Fig.~\ref{fig:comparison_fixed_vs_varying_cosmo}.
These results are compared with those obtained by fixing
the 6 $\Lambda$CDM cosmological parameters in Fig.~\ref{fig:comparison_fixed_vs_varying_cosmo}. Given the good agreement that we find for this LN template, we proceed by keeping fixed the 6 $\Lambda$CDM cosmological parameters to reduce the computational cost of the MCMC analysis.

\begin{figure}
\centering
	\includegraphics[width=0.45\columnwidth]{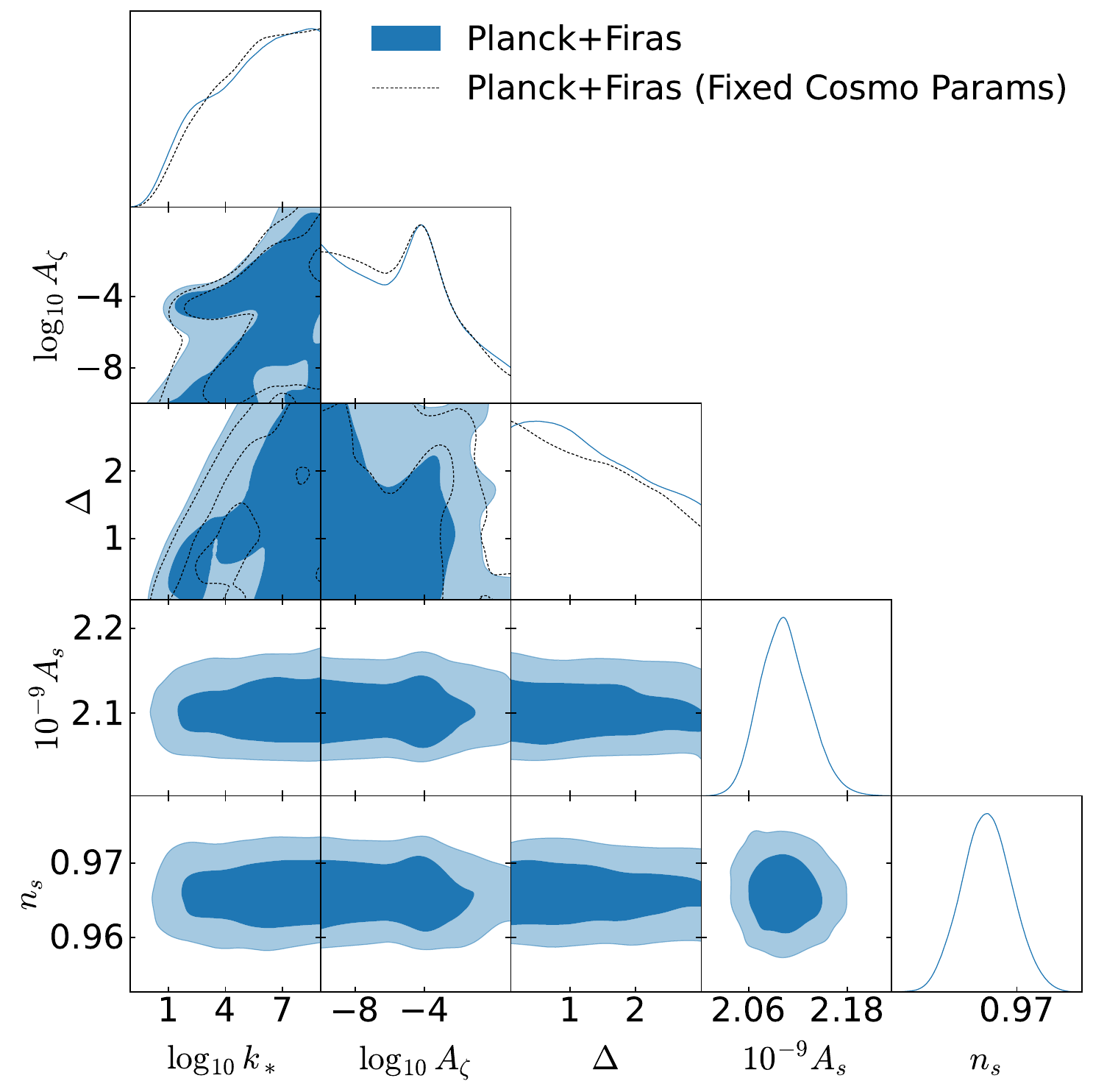}
	\includegraphics[width=0.45\columnwidth]{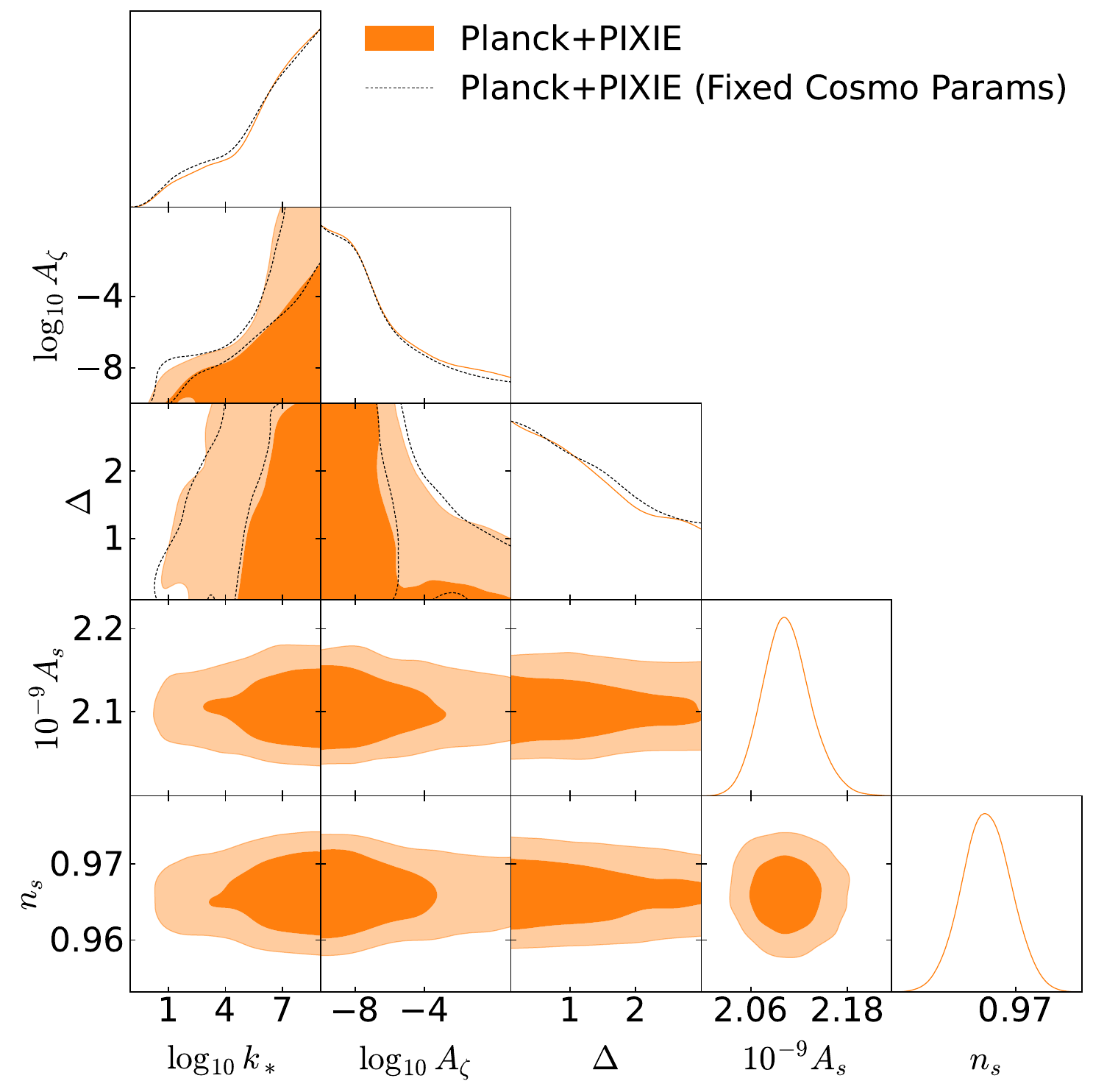}
	\caption{Comparison of the triangle plots obtained in the \emph{Case 1} for both a fixed and a varying $\Lambda$CDM cosmology. For simplicity, we restrict the priors on $\log_{10}A_\zeta$ and $\log_{10}k_* [\si{\per\mega\parsec}]$ within the ranges $[-10,-1]$ and $[-1,9]$ respectively.}
\label{fig:comparison_fixed_vs_varying_cosmo}
\end{figure}

In order to forecast the impact of future CMB SD measurements on the scalar-induced interpretation of PTA data, we compute the 3d marginalized posterior distribution of $(\log_{10} A_\zeta, \log_{10} k_*, \Delta)$ from the public chains of the NG15 analysis obtained by using the SIWGB associated to the LN model in  Eq.~1 in the main text. 
We then add the log-value of such normalized posterior to the total CMB log-likelihood $\ln\mathcal{L}_{\rm Planck} + \ln \mathcal{L}_{\rm SD}$.

\begin{figure*}
	\centering
	\includegraphics[width=0.45\columnwidth]{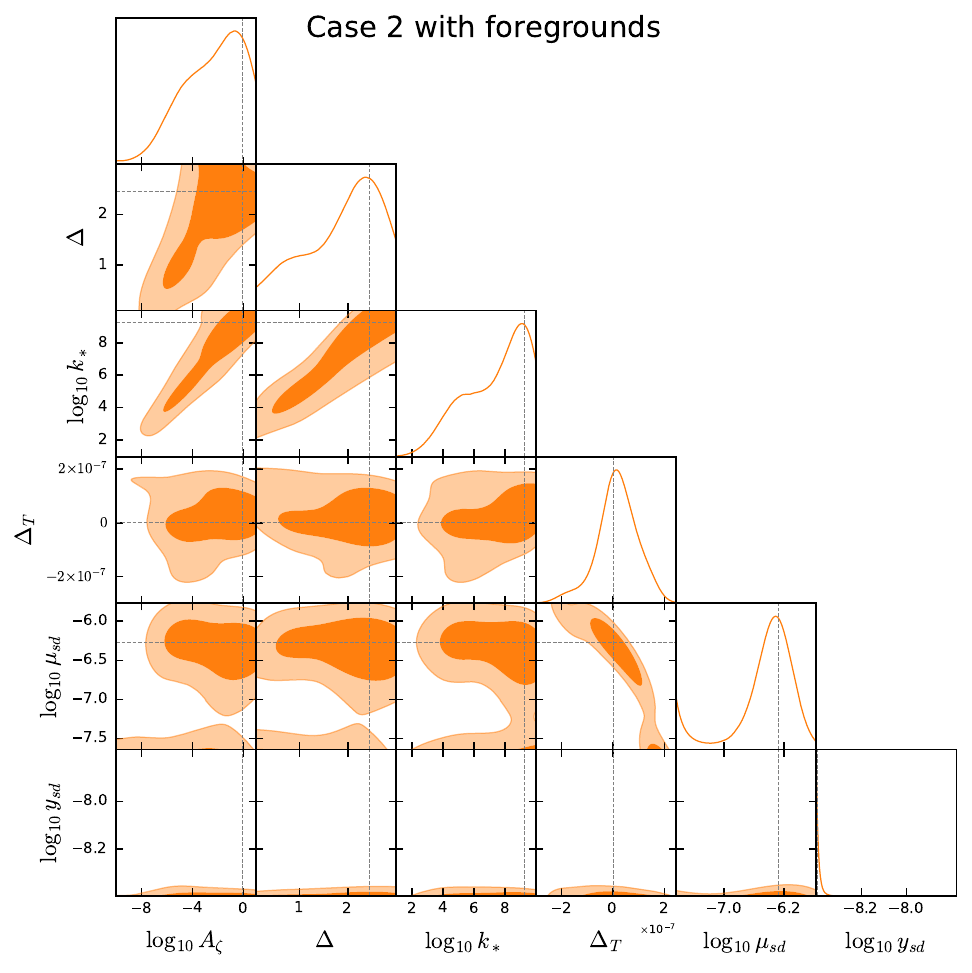}
    \hspace{-0.35cm}
	\includegraphics[width=0.45\columnwidth]{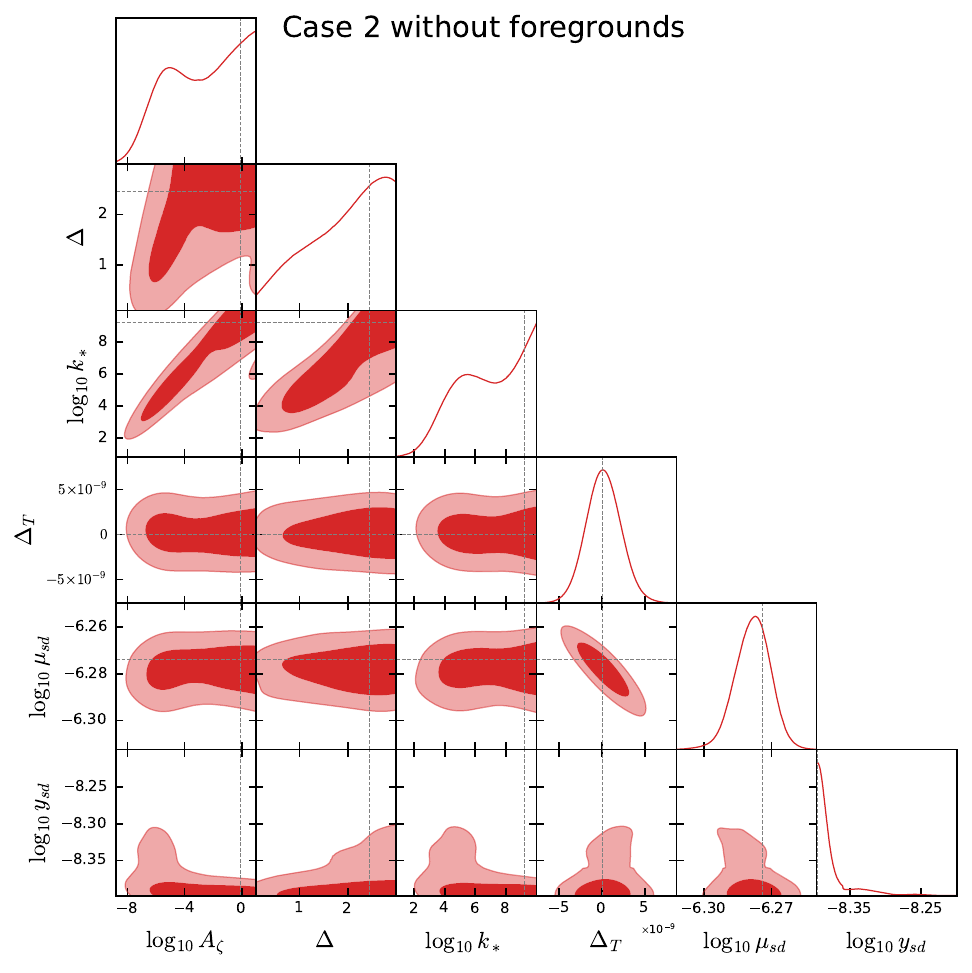}
	\includegraphics[width=0.91\columnwidth]{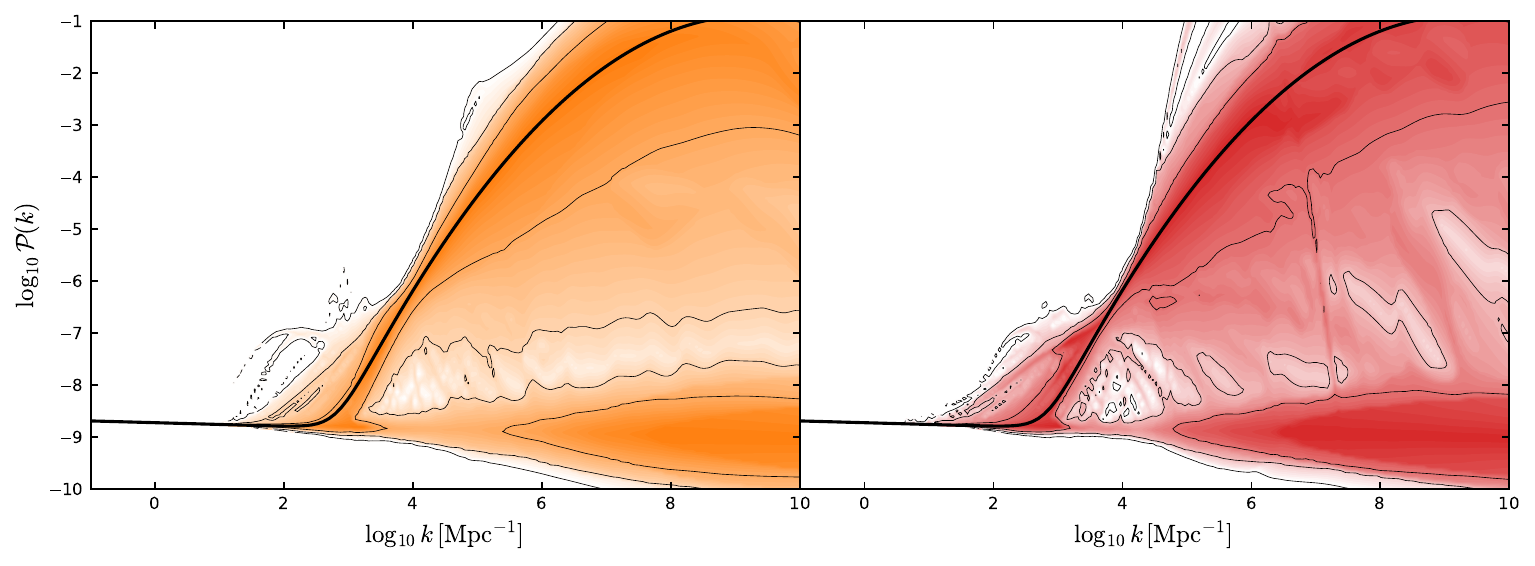}
    \includegraphics[width=0.91\columnwidth]{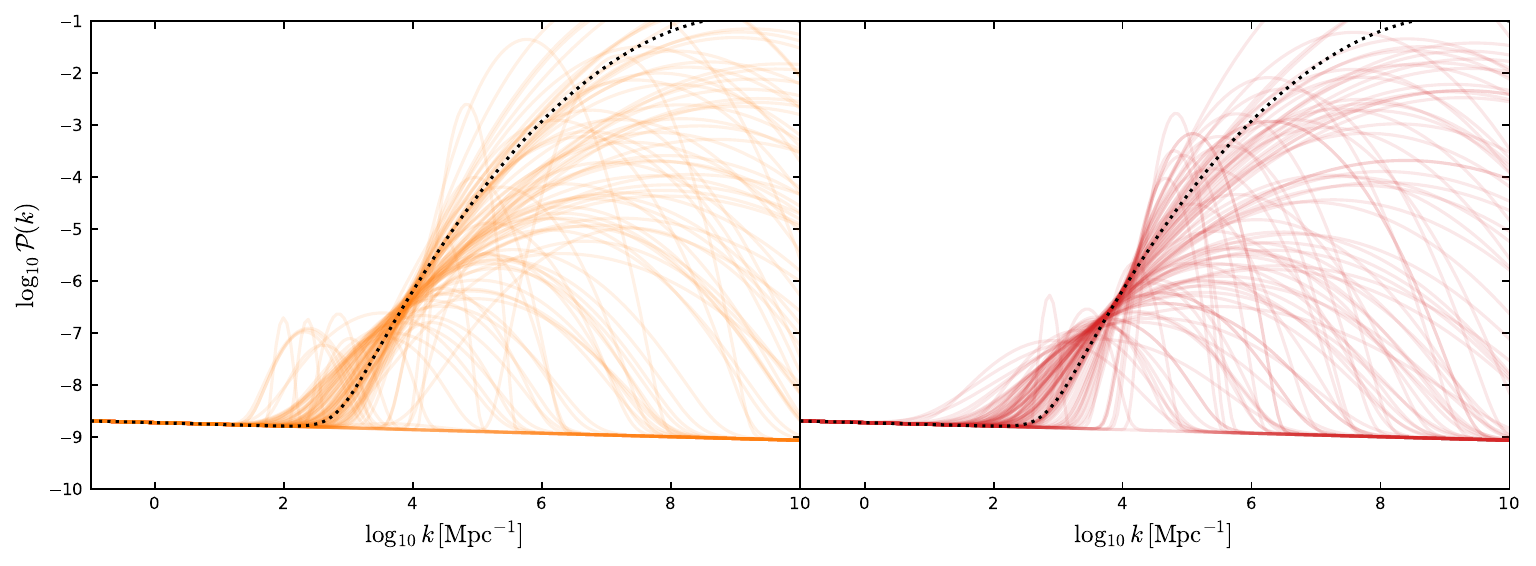}
	\caption{[Top] triangle plots, [center] PPD for $\mathcal{P}(k)$, [bottom] and posterior samples for \emph{Case 2} varying (left) and keeping the foregrounds fixed to their values (right). The fiducial $\mathcal{P}(k)$ is plotted with a black dotted line. The color shades of the PPD range from $0$ (darker shades) to $3\,\sigma$ (lighter shades).}
	\label{fig:case2-fg-nofg}
\end{figure*}

\medskip
\noindent
{\bf The role of SD likelihoods in the PPS reconstruction.}
We now discuss the relevance of SD likelihoods for the reconstruction of the PPS.  
In the top panels of Fig.~\ref{fig:case2-fg-nofg}, we add the derived posteriors on $\mu,\,y$ and $\Delta_T$ to the PIXIE forecast shown in Fig.~2 in the main text. Fig.~\ref{fig:case2-fg-nofg} shows how the constraints on  $\mu$ are significantly affected by foregrounds, consistently with~\cite{Schoneberg:2020nyg}. 
By fixing the foregrounds, we get a significant detection $\mu =  \left(\,5.3\pm 0.1\,\right)\times 10^{-7}$ at 68\% CL.
On the other hand, when we vary the foreground parameters in the analysis and marginalize over them, we only obtain a 95\% CL upper bound on $\mu$, i.e. $8.5\times 10^{-7}$. 

The role of foregrounds is instead different for the inference of the PPS by CMB SDs.
Indeed, degeneracies among the LN parameters cannot be broken, even when foregrounds are removed and $\mu$ is detected with high statistical significance.
This fact is easy to understand by looking at Eq.~\eqref{eq:sds-from-pps}. 
Indeed, the detection of $\mu$ we can only constrain a specific combination of LN parameters, but not each of them individually.
This clarifies why the PPD of $\mathcal{P}(k)$, despite hinting at a strong scale dependence at $\sim1\sigma$ in the region $k\in[10^2,\,10^5]\,{\rm Mpc}^{-1}$ (see the dark-shaded areas in Fig.~\ref{fig:case2-fg-nofg}), is consistent with a near scale invariant spectrum.
To clarify the PPD, we also plot the posterior samples in the bottom panels of the Fig.~\ref{fig:case2-fg-nofg}. 
Both small peaks in the SD window and larger ones at larger wavenumbers are admitted by the MCMC, as they yield the same value of $\mu$.

Measuring $y$ generated by the dissipation of the acoustic waves would break the degeneracies mentioned above, and assess the primordial origin of the SDs. 
However, $y$ generated by the dissipation of the acoustic waves in \emph{Case 2} is too small to be detected, regardless of the way we treat foregrounds in our forecast, see Fig.~1 in the main text. 
This example can be compared to our forecast for {\em Case 3} where $y$ is instead large enough to be detected by PIXIE, resulting in $\left(\,4.9^{+4.7}_{-4.3}\,\right)\times 10^{-8}$ at 95\% CL. In that case, as shown in Fig.~2 in the main text, the PPS parameters can be all constrained, and the $\mathcal{P}(k)$ can be reconstructed very well.

\end{widetext}

\end{document}